\documentclass[preprint,12pt]{elsarticle}
\usepackage{epsfig}
\usepackage{graphicx}
\usepackage{amssymb}

\journal{Physics Letters B}

\begin{document}

\begin{frontmatter}

\title{First measurement of the circular beam asymmetry in the $\vec \gamma p \to \pi^0\eta p$
reaction}

\cortext[cor1]{Corresponding author}
\author[1,2]{V.~L.~Kashevarov}\corref{cor1}\ead{kashev@kph.uni-mainz.de}
\author[3]{A.~Fix}
\author[1]{P.~Aguar-Bartolom\'e}
\author[1]{L.~K.~Akasoy}
\author[4]{J.~R.~M.~Annand}
\author[1]{H.~J.~Arends}
\author[5]{K.~Bantawa}
\author[6]{R.~Beck}
\author[7]{V.~Bekrenev}
\author[8]{H.~Bergh\"auser}
\author[9]{A.~Braghieri}
\author[10]{D.~Branford}
\author[11]{W.~J.~Briscoe}
\author[12]{J.~Brudvik}
\author[2]{S.~Cherepnya}
\author[4]{R.~F.~B.~Codling}
\author[11]{B.~T.~Demissie}
\author[1,4]{E.~J.~Downie}
\author[8]{P.~Drexler}
\author[2]{L.~V.~Fil'kov}
\author[10]{D.~I.~Glazier}
\author[8]{R.~Gregor}
\author[4]{D.~Hamilton}
\author[1,11]{E.~Heid}
\author[13]{D.~Hornidge}
\author[14]{I.~Jaegle}
\author[1]{O.~Jahn}
\author[10]{T.~C.~Jude}
\author[4]{J.~D.~Kellie}
\author[14]{I.~Keshelashvili}
\author[15]{R.~Kondratiev}
\author[16]{M.~Korolija}
\author[8]{M.~Kotulla}
\author[7]{A.~Koulbardis}
\author[7]{S.~Kruglov}
\author[14]{B.~Krusche}
\author[15]{V.~Lisin}
\author[4]{K.~Livingston}
\author[4]{I.~J.~D.~MacGregor}
\author[14]{Y.~Maghrbi}
\author[5]{D.~M.~Manley}
\author[1]{M.~Martinez-Fabregate}
\author[4]{J.~C.~McGeorge}
\author[4]{E.~F.~McNicoll}
\author[16]{D.~Mekterovic}
\author[8]{V.~Metag}
\author[16]{S.~Micanovic}
\author[13]{D.~Middleton}
\author[9]{A.~Mushkarenkov}
\author[12]{B.~M.~K.~Nefkens}
\author[6]{A.~Nikolaev}
\author[8]{R.~Novotny}
\author[1]{M.~Ostrick}
\author[1]{P. B.~Otte}
\author[1,11]{B.~Oussena}
\author[9]{P.~Pedroni}
\author[14]{F.~Pheron}
\author[15]{A.~Polonski}
\author[12]{S.~N.~Prakhov}
\author[4]{J.~Robinson}
\author[4]{G.~Rosner}
\author[14]{T.~Rostomyan}
\author[1]{S.~Schumann}
\author[10]{M.~H.~Sikora}
\author[17]{D.~Sober}
\author[12]{A.~Starostin}
\author[11]{I.~I.~Stakovsky}
\author[12]{I.~M.~Suarez}
\author[16]{I.~Supek}
\author[10]{C.~Tarbert}
\author[8]{M.~Thiel}
\author[1]{A.~Thomas}
\author[1]{M.~Unverzagt}
\author[10]{D.~P.~Watts}
\author[14]{D.~Werthm\"uller}
\author[16]{I.~Zamboni}
\author[14]{and~F.~Zehr}

\author{(The~Crystal~Ball~at~MAMI,~TAPS,~and~A2~Collaborations)}

\address[1]{Institut f\"ur Kernphysik, Johannes Gutenberg-Universit\"at Mainz, Mainz, Germany}
\address[2]{Lebedev Physical Institute, Moscow, Russia}
\address[3]{Tomsk Polytechnic University, Tomsk, Russia}
\address[4]{Department of Physics and Astronomy, University of Glasgow, Glasgow, UK}
\address[5]{Kent State University, Kent, OH, USA}
\address[6]{Helmholtz-Institut f\"ur Strahlen- und Kernphysik, Universit\"at Bonn, Bonn, Germany}
\address[7]{Petersburg Nuclear Physics Institute, Gatchina, Russia}
\address[8]{II.\ Physikalisches Institut, Universit\"at Giessen, Giessen, Germany}
\address[9]{INFN Sezione di Pavia, Pavia, Italy}
\address[10]{School of Physics, University of Edinburgh, Edinburgh, UK}
\address[11]{The George Washington University, Washington, DC, USA}
\address[12]{University of California at Los Angeles, Los Angeles, CA, USA}
\address[13]{Mount Allison University, Sackville, NB, Canada}
\address[14]{Institut f\"ur Physik, Universit\"at Basel, Basel, Switzerland}
\address[15]{Institute for Nuclear Research, Moscow, Russia}
\address[16]{Rudjer Boskovic Institute, Zagreb, Croatia}
\address[17]{The Catholic University of America, Washington, DC, USA}

\begin{abstract}
The circular photon asymmetry for $\pi^0\eta$ photoproduction on the proton was measured for
the first time at the tagged photon facility of the MAMI C accelerator
using the Crystal Ball/TAPS photon spectrometer.
The experimental results are interpreted within a phenomenological isobar model
that confirms the dominant role of the $\Delta(1700)D_{33}$ resonance. The
measured asymmetry allows us to identify small contributions from
positive-parity resonances via interference terms with the dominant $D_{33}$ amplitude.
\end{abstract}


\end{frontmatter}

The full understanding of the structure and excitation spectrum of the nucleon
remains one of the most challenging topics of particle physics.
Traditionally, a large amount of information on
baryon resonances was provided by scattering or photoproduction of pions. However, despite
manifest progress, our knowledge of the properties of many resonances is still rather limited,
to the extent that the very existence of some so-called established states \cite{PDG}
is called into question.
The production of meson pairs like $\pi\pi$ or $\eta\pi$ is an attractive
tool to study resonances that couple strongly to intermediate $\Delta \pi, \Delta \eta$ or
$N^* \pi$ states.

In recent years, the empirical information available on  the $\gamma p\to \pi^0\eta
p$ reaction has considerably improved. New data were obtained on the total and differential cross
sections  as well as on linear beam asymmetries
\cite{Horn,Weinheimer,Tohoku,Ajaka,Kashev,GutzBeam}.
The observed rapid rise of the total cross section from threshold together with
an almost isotropic $\eta$ angular distribution
already indicates that the $\eta$ meson is mostly emitted into an $s$-wave state with respect to
the $\pi^0p$ system. Considering that the $\pi^0p$ interaction is mainly due to excitation of
the $\Delta(1232)$, this points to the importance of a partial-wave amplitude with spin-parity
$J^P = 3/2^-$ and isospin $I = 3/2$. In $\pi$N scattering these quantum numbers require a
$d$-wave and the amplitude is therefore called $D_{33}$. It
is populated by the $\Delta(1700)D_{33}$ resonance in the energy region close
to the threshold of the $\gamma p\to \pi^0 \eta p$ reaction.
Furthermore, within phenomenological models \cite{Horn,Dor,FKLO,FOT}
a reasonable description of the total and differential cross sections
as well as  the measured beam asymmetries \cite{Ajaka,GutzBeam} can be achieved
assuming the dominance of this amplitude.
However, the reliable extraction of resonance parameters and the study of
small contributions from other partial-wave amplitudes require
measuring additional spin observables. The
nontrivial role of such resonances was indicated by the PWA analysis of ref.\,\cite{Horn}
as well as by measurements of the angular distributions presented in
refs.\,\cite{Kashev,FKLO}.

The general structure of the cross section for the photoproduction of
two pseudoscalar mesons and the definition of spin observables is discussed
in ref.\,\cite{Roberts}.
The circular beam asymmetry $I^{\odot}$ has been measured in double pion
photoproduction \cite{Strauch, Kram}. In the case of $\pi\eta$ photoproduction
related observables, $I^s$ and $I^c$, have recently been measured with a linearly polarized
beam \cite{Gutz}. Model calculations \cite{Roca,Doring}
clearly demonstrate the strong sensitivity of these spin observables to the
dynamical content of the reaction amplitude.

In this Letter we present first measurements of the circular beam asymmetry
in the reaction $\vec{\gamma} p\to\pi^0\eta p$.
The experiment was performed with the Crystal Ball/TAPS hermetic spectrometer system
at the Glasgow tagged photon facility \cite{TAGGER} of the MAMI C accelerator in Mainz \cite{MAMI}.
The experimental setup and the event selection procedure are
described in detail in \cite{Kashev}.
The data were taken in April 2009 (300 hours with a 10 cm long
liquid hydrogen target and a beam current of 10 nA).

\begin{figure}
\begin{center}
\resizebox{0.7\textwidth}{!}{%
\includegraphics{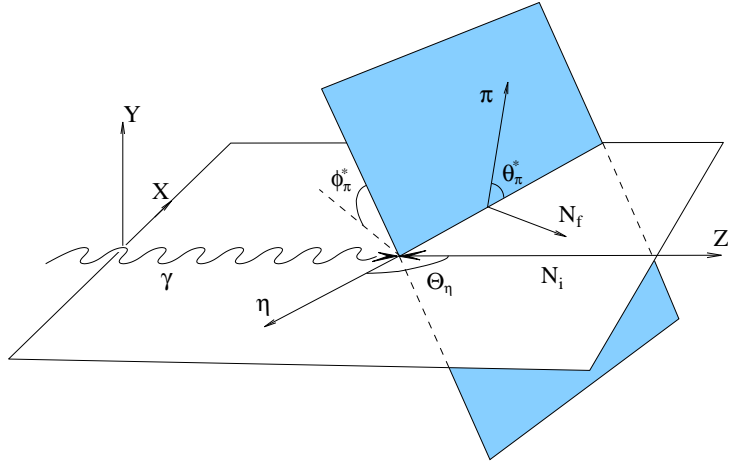}}
\caption{Diagrams representing the angles $\Theta_\eta$ and $\Omega=(\theta,\phi)$ used
for description of the reaction kinematics.} \label{fig1}
\end{center}
\end{figure}

The usual circular photon asymmetry (in the literature also called the helicity photon
asymmetry) is defined as
\begin{equation} \label{10}
I^{\odot}(\phi)=\frac{1}{P_\gamma}\,\frac{d\sigma^+-d\sigma^-}{d\sigma^++d\sigma^-}\,,
\end{equation}
where $d\sigma^+(d\sigma^-)$ denote the
5-fold differential cross sections integrated over the energy
of the $\eta$ meson, its solid angle,
$d\Omega_\eta$, and the polar angle of the pion, $d\Theta_\pi$,
for beam helicities $\lambda_\gamma=\pm 1$. $P_\gamma$ is the degree of circular beam
polarization.
The argument $\phi$ is the angle between the reaction plane and the plane spanned by the
momenta of the produced pion and the proton in the final state (see Fig.\,\ref{fig1}).
It is equal to
the pion azimuthal angle in the $\pi N$ c.m.\ frame with the $z$-axis being in the
opposite direction to the $\eta$ momentum. In the case of a two-body final state,
the asymmetry $I^{\odot}$
vanishes exactly.

In what follows, we will consider an observable whose definition slightly differs from
(\ref{10}), namely
\begin{equation} \label{10a}
W^c(\phi)=\frac{2\pi}{\sigma}\ I^{\odot}(\phi)\frac{d\sigma}{d\phi}=
\frac{1}{P_\gamma}\,\frac{\pi}{\sigma}\frac{d\sigma^+-d\sigma^-}{d\phi}\,,
\end{equation}
where the unpolarized cross section is defined as
\begin{equation} \label{10b}
\frac{d\sigma}{d\phi}=\frac{1}{2}\frac{d\sigma^++d\sigma^-}{d\phi}\,
\end{equation}
and $\sigma$ stands for the integrated (total) cross section,
\begin{equation}\label{10c}
\sigma=\int_0^{2\pi}\frac{d\sigma}{d\phi}\ d\phi\,.
\end{equation}
As can readily be seen from (\ref{10}) and (\ref{10a}), the only difference between the
quantity $W^c$ and the asymmetry $I^{\odot}$ is that the former contains in the
denominator not the differential cross section at a given angle $\phi$ but its average
over the whole $\phi$ region, which is equal to the total cross section $\sigma$ divided
by $2\pi$. Contrary to
$I^{\odot}$, the function $W^c$ does not contain any $\phi$-dependence in the
denominator, thus the most interesting part, the angular dependence of the helicity
difference $d\sigma^+ - d\sigma^-$, is directly visible.

The asymmetry $W^c$ is determined from the experimental data, using the number of
reconstructed events for each helicity
corrected for the detector acceptance and photon flux. All other normalization factors cancel
in the ratio (\ref{10a}). The degree of polarization, $P_\gamma$, is given by the
electron beam polarization multiplied by a well known factor describing the helicity transfer
to photons in  bremsstrahlung processes \cite{Olsen} which varies between $84\%$ and $99\%$
in our photon energy range.
The electron beam polarization was measured by Mott and Moeller scattering to be
$P_e = (80.9\pm 2.5)\%$.

\begin{figure}
\begin{center}
\resizebox{0.8\textwidth}{!}{%
\includegraphics{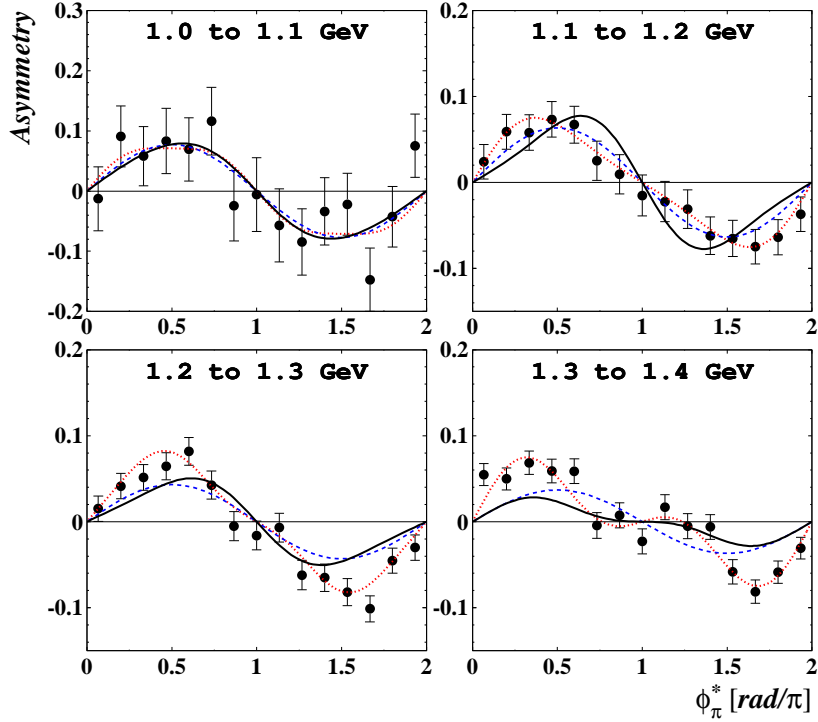}}
\caption{Angular distributions of the cross-section asymmetry $W^c(\phi)$ for the
reaction $\vec \gamma p\to\pi^0\eta p$ as determined according to eq.\,(\ref{10a}).
The dotted curve is our
fit with three terms included in the Fourier expansion \protect(\ref{13}). The solid curves
show predictions
of the full isobar model with six resonances whose parameters were fitted to the angular
distributions for $\gamma p\to\pi^0\eta p$ as described in ref.\,\cite{FKLO}. The dashed
curve is a similar prediction with only the $D_{33}$ amplitude taken into account.} \label{fig2}
\end{center}
\end{figure}

The results for the observable $W^c$ are shown
in Fig.\,\ref{fig2} for different beam energy bins with statistical uncertainties only.
The systematic uncertainty is dominated by the contribution from the beam polarization.
Further contributions from acceptance calculations and flux normalization are
negligible.

Due to parity conservation the cross sections for different photon helicities are related by
\begin{equation}\label{12a}
d\sigma^+(-\phi)=d\sigma^-(\phi)\,.
\end{equation}
Then the relations
\begin{equation}\label{12b}
W^c(2\pi-\phi)=W^c(-\phi)=-W^c(\phi)
\end{equation}
immediately follow from the definition in eq.(\ref{10a}). As a consequence, $W^c(\phi)$ may be
Fourier-expanded over the functions $\sin n\phi$,
\begin{equation}\label{13}
W^c(\phi)=\sum_{n=1}^{n_{\rm max}} A_n\,\sin n\phi\,.
\end{equation}
In the absence of a strong background, the maximum value
of $n$ is determined by the maximum spin of the contributing resonances.
The same features were demonstrated for $2\pi$ photoproduction in
refs.\,\cite{Strauch,Kram}.
While in the channels with charged pions ($\pi^+\pi^-$, $\pi^0\pi^+$) the asymmetry $I^{\odot}(\phi)$
exhibits quite a complicated angular dependence, due to strong background contributions,
the $\pi^0\pi^0$ production is easily fitted by $\sin 2\phi$.
As in $\pi^0 \eta$, the production of two neutral pions is dominated by
resonance excitations, whereas, in the charge channels, different peripheral background
mechanisms become important. These are responsible for the rather complex angular dependence
of the helicity asymmetry.

\begin{figure}
\begin{center}
\resizebox{0.9\textwidth}{!}{%
\includegraphics{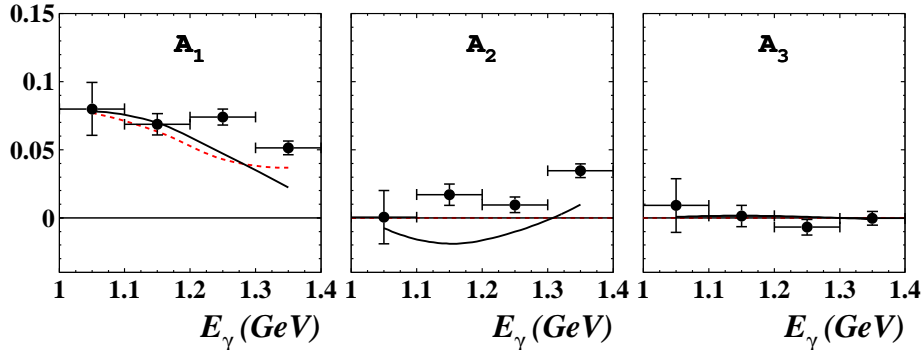}}
\caption{Coefficients $A_n$ $(n=1,2,3)$ of the $\sin n\phi$ expansion of the asymmetry
$W^c(\phi)$, Eq. (\ref{13}). The solid line is the full model prediction, the dashed 
line includes only the $D_{33}$ amplitude.} 
\label{fig3}
\end{center}
\end{figure}

To fit our data for $W^c(\phi)$ we retain the first three terms in eq.\,(\ref{13}). The
results are presented in Fig.\,\ref{fig2} by the dotted line. In Fig.\,\ref{fig3} the
first three coefficients $A_1$, $A_2$, and $A_3$ from the series (\ref{13}) are plotted
as functions of the photon energy. The first term in the series
is most important. The terms with $n=2$ and $n=3$ are significantly suppressed.

In the same figure we show model predictions obtained with the formalism developed in
refs.\,\cite{FKLO} and all parameters fixed by fitting the angular distributions with
only a $D_{33}$ amplitude (dashed) and the full isobar model (solid). Here we limit
ourselves to a brief overview of the formal basis. The general structure of the matrix
element is represented by a background amplitude  $t^B_{m_f\lambda}$ and a resonance part
$t^R_{m_f\lambda}$:
\begin{equation}\label{15}
t_{m_f\lambda}=t^B_{m_f\lambda}+\sum_{R(J^\pi;T)}t^R_{m_f\lambda}\,,
\end{equation}
where the summation is over the resonance states $R(J^\pi;T)$ determined by their
spin-parity $J^\pi$ and the isospin $T$. The indices $m_f=\pm 1/2$ and $\lambda=\pm 1/2$,
$\pm 3/2$ denote respectively the $z$-projection of the final nucleon spin and the
initial state helicity.

In the isobar model the transition to the three particle state $\pi\eta N$ is described
in terms of intermediate decays into the $\eta + P_{33}(1232)$ and $\pi+S_{11}(1535)$
states, followed by the decay of the $P_{33}(S_{11})$ resonance into $\pi N(\eta N)$. The
resonance amplitudes are accordingly decomposed into two parts
\begin{equation}\label{15a}
t^R_{m_f\lambda}=t^{R(\eta\Delta)}_{m_f\lambda}+t^{R(\pi N^*)}_{m_f\lambda}\,,
\end{equation}
where we use the notations $\Delta$ and $N^*$ for $P_{33}(1232)$ and $S_{11}(1535)$,
respectively.

The polarized cross section is expressed as a difference between
quadratic forms of the amplitudes (\ref{15}) with different helicities:
\begin{equation}\label{35}
I^{\odot}\frac{d\sigma}{d\phi}\sim\sum_{m_f,\,\lambda=\frac12,\frac32}\int\left(\left|\,t_{m_f\lambda}\right|^2
-\left|\,t_{m_f-\lambda}\right|^2\right)d\,\mbox{Ps}\,,
\end{equation}
$d$\,Ps is the appropriate element of the phase space.

As is shown in \cite{FKLO,FOT} the background is small, so that we neglect its contribution
and assume that in our energy region the reaction is dominated by
the $D_{33}$ amplitude accompanied by a relatively small admixture of other partial waves,
in our case $P_{31}$, $P_{33}$, and $F_{35}$ which contribute via
interference with the dominant $D_{33}$ excitation.
Keeping only interference terms which are
linear in the `weak' amplitudes, the integrand in eq.\,(\ref{35}) calculated to
the first order in $t^{P_{31}}$, $t^{P_{33}}$, and $t^{F_{35}}$ reads
\begin{eqnarray}\label{40}
&&\left|\,t_{m_f\lambda}\right|^2 -\left|\,t_{m_f-\lambda}\right|^2\simeq
\left|\,t^{D_{33}}_{m_f\lambda}\right|^2+2\Re e\Big\{\,
\bar{t}_{m_f\lambda}^{\,D_{33}}t_{m_f\lambda}^{P_{31}} \nonumber\\
&&\phantom{xxx}+ \bar{t}_{m_f\lambda}^{\,D_{33}}t_{m_f\lambda}^{P_{33}}+
\bar{t}_{m_f\lambda}^{\,D_{33}}t_{m_f\lambda}^{F_{35}}\Big\}-(\lambda\to -\lambda),
\end{eqnarray}
where $\bar{t}$ means the complex conjugated value. Using eq.\,(\ref{40}) one obtains for
the asymmetry~(\ref{10a}) the following expression \cite{FixAr}:
\begin{equation}\label{45}
W^c(\phi)=A_1\sin\phi+A_2\sin2\phi\,.
\end{equation}

Explicit expressions for the coefficients $A_n$ are not important for the discussion below.
They will be presented in a separate paper \cite{FixAr}. Here we
simply summarize the key points to demonstrate the main physical ideas.
The first coefficient in eq.\,(\ref{45}), $A_1$, is determined solely by the
$D_{33}$ wave. Actually, it also contains terms that are
quadratic in `small' partial-wave amplitudes. However, as noted above, the latter were
neglected here due to their insignificance. It is also worth noting that $D_{33}$
contributes to the asymmetry $W^c(\phi)$ only because the decay modes,
$D_{33}\to\pi N^*$ and $D_{33}\to\eta\Delta$, interfere with each other.
In \cite{Kashev} the contribution of the $D_{33}\to\pi N^*$ decay
was identified using the characteristic form of the pion
angular distribution in the $\pi N$ c.m.\ system. The present results for $W^c(\phi)$
are additional independent evidence for the importance of this decay channel.

If only the linear contribution of `weak' resonance amplitudes is retained they enter
only the second term of eq.\,(\ref{45}).
In other words, the $\sin 2\phi$ admixture in the asymmetry $W^c$
appears only through the interference of these states with the dominant $D_{33}$ wave.
It can be shown \cite{FixAr} that the appearance of this term in the expansion (\ref{45}) is
caused by the different parity of the resonances $P_{31}$, $P_{33}$, and
$F_{35}$ relative to the dominant $D_{33}$ state. In this respect, the term with $\sin
2\phi$ is a signature of positive parity states entering the amplitude.

The solid curves in these figures show predictions of the full isobar model including
the $D_{33}(1700)$, $P_{33}(1600)$, $P_{31}(1750)$, $F_{35}(1905)$, $P_{33}(1920)$, and
the $D_{33}(1940)$. The dashed curve includes only the $D_{33}$ amplitude. 
The resonance parameters were taken from ref.\,\cite{FKLO} where they were obtained by fitting
the measured angular distributions \cite{Kashev}.
We use the parameter set, corresponding to Solution I compiled in Table II of ref.\,\cite{FKLO} 
which also provides reasonable agreement with the measured beam asymmetry \cite{Ajaka,Gutz}. 
The present data were not used in the fit. In view of strong sensitivity of $W^c$ to the 
resonance content of the amplitude, in particular to the  interference effects between different 
partial waves, the quality of description in Figs.\,\ref{fig2} and \ref{fig3} is rather good. 
Nevertheless, systematic deviations at higher energies point to necessity for further improvements 
of the theory. As one can see, at $E_\gamma>1.2$\,GeV our calculation underestimates the 
coefficient $A_1$. In this region the single $D_{33}$ model (only the $D_{33}$ wave is included) 
provides even a slightly better description of the asymmetry.
The coefficient $A_2$ is predicted to be negative in contrast to the experimental value, although 
the general tendency of its behavior, the monotonic increase with the energy, is roughly 
reproduced. The coefficient $A_3$, proportional to squares of the `weak' amplitudes, is 
comparable with zero in the whole energy region, in accordance with the data.

In summary, we presented first measurements of the circular photon asymmetry in the
reaction $\vec \gamma p\to\pi^0\eta p$. Our purpose was to explore the contribution
of resonances with positive parity, which manifest themselves primarily in
polarization observables via interference with the dominating $D_{33}$ partial wave.
The experimental results were compared to the model predictions with the parameters
fixed in ref.\,\cite{FKLO} by fitting the unpolarized angular distributions of ref.\,\cite{Kashev}.
The comparison demonstrates that the single $D_{33}$ model roughly reproduces the gross
features of the observed helicity asymmetry. At the same time, in the region $E_\gamma>1.3$\,GeV 
the positive parity resonances start to come into play, resulting in strong increase of 
the $\sin 2\phi$ term in the Fourier expansion (\ref{13}). The full isobar model of 
ref.\,\cite{FKLO} including $P_{33}(1600)$, $P_{31}(1750)$, $F_{35}(1905)$, and $P_{33}(1920)$ 
states demonstrates certain shortcomings, especially at higher energies. The present measurements 
together with the linear beam asymmetry presented in ref.\,\cite{GutzBeam} open a path for 
further improvements of the theoretical description of $\pi^0\eta$ photoproduction.

The authors wish to acknowledge the excellent support of the accelerator group and
operators of MAMI. This work was supported by the Deutsche Forschungsgemeinschaft (SFB
443, SFB/TR16), DFG-RFBR (Grant No. 09-02-91330), the European Community-Research
Infrastructure Activity under the FP6 ``Structuring the European Research Area''
programme (Hadron Physics, contract number RII3-CT-2004-506078), Schweizerischer National
fonds, the UK Engineering and Physical Sciences Research Council, U.S. DOE, U.S. NSF, and
NSERC (Canada). A.F. acknowledges additional support from the RF Federal programm ``Kadry''.
We thank the undergraduate students of Mount Allison University and
The George Washington University for their assistance.


\end{document}